\documentclass[aps,prx,superscriptaddress,twocolumn,nofootinbib,nobibnotes,floatfix,showpacs,reprint,longbibliography]{revtex4-2}
\usepackage{graphicx}
\graphicspath{{Figures/}}
\usepackage[table]{xcolor}
\usepackage{ragged2e}
\usepackage{booktabs}
\usepackage{siunitx}
\usepackage[normalem]{ulem}
\usepackage{xr-hyper}
\usepackage{braket}

\usepackage{hyperref}
\usepackage{cleveref}

\begin{document}

\bibliographystyle{unsrt}

\title{Symmetry-Protected Phonon Topology and Low Lattice Thermal Conductivity in Square–Octagonal Chalcogenides}
\author{Surabhi Suresh Nair}
\affiliation{Department of Physics, Khalifa University, Abu Dhabi-127788, United Arab Emirates (UAE)}
\author{Chiranjit Mondal}
\affiliation{Department of Physics and Astronomy, Seoul National University, Seoul 08826, Korea}
\affiliation{Center for Theoretical Physics (CTP), Seoul National University, Seoul 08826, Korea}
\author{Aftab Alam}
\email{aftab@iitb.ac.in}
\affiliation{Department of Physics, Indian Institute of Technology, Bombay, Powai, Mumbai 400076, India}
\author{Nirpendra Singh}
\email{Nirpendra.Singh@ku.ac.ae}
\affiliation{Department of Physics, Khalifa University, Abu Dhabi-127788, United Arab Emirates (UAE)}
\affiliation{Research and Innovation Center for Graphene and 2D materials (RIC2D), Khalifa University, Abu Dhabi, United Arab Emirates}

\begin{abstract}
Unconventional lattice geometries provide an effective platform for realizing symmetry-protected topological phonon states that can strongly influence lattice thermal transport. In this work, we explore the relationship between topological phonon band features and thermal transport in square--octagonal ($so$) chalcogenide monolayers, namely MoS$_2$ and SnS, by combining first-principles calculations with phonon Boltzmann transport theory. Symmetry analysis reveals the presence of nontrivial phonon band topology in the form of symmetry-protected nodal lines. Crossings between nodal lines carrying different symmetry eigenvalues produce fourfold Dirac points that enhance the phonon group velocity ($v_g$), whereas nearly flat nodal lines lead to strong suppression of $v_g$. The coexistence of these features, together with substantial phonon softening and enhanced anharmonic scattering around the topological band crossings, markedly suppresses the lattice thermal conductivity ($\kappa_l$). As a result, room-temperature $\kappa_l$ values of 4.0 W/mK for $so$-SnS and 18.7 W/mK for $so$-MoS$_2$ are obtained, representing reductions by more than a factor of two and eight, respectively, relative to their hexagonal phases. Our results uncover a direct connection between phonon band topology and heat transport in two-dimensional materials, highlighting lattice symmetry and topological band engineering as promising routes for tailoring thermal properties. These findings further suggest opportunities for designing topological phononic and thermoelectric devices with controllable heat flow.
\end{abstract}
\maketitle

\section{Introduction}
The continuous loss of waste heat from industrial operations and automobile exhaust systems remains a major challenge for sustainable energy utilization. Thermoelectric (TE) materials offer a direct and reversible approach for converting heat into electricity, enabling energy recovery in applications ranging from wearable electronics to space technologies. However, many efficient TE compounds contain toxic or rare elements, thereby motivating the search for environmentally friendly, earth-abundant alternatives.
The efficiency of a thermoelectric material is characterized by the dimensionless figure of merit, $ZT = \frac{S^2 \sigma T}{\kappa_e + \kappa_l},$
where $S$ denotes the Seebeck coefficient, $\sigma$ the electrical conductivity, $T$ the absolute temperature, and $\kappa_e$ and $\kappa_l$ represent the electronic and lattice components of the thermal conductivity, respectively. Although $S$, $\sigma$, and $\kappa_e$ are strongly interdependent, the lattice thermal conductivity $\kappa_l$, governed by phonon transport, can be manipulated more independently and therefore plays a central role in improving thermoelectric performance, particularly in semiconducting systems.

Phonon transport is highly sensitive to lattice symmetry, geometric arrangement, and chemical bonding. Crystal symmetry determines phonon degeneracies and scattering channels: preserving symmetry maintains degeneracies, whereas symmetry breaking lifts them, enlarges the phonon scattering phase space, and suppresses $\kappa_l$. For example, strain-driven symmetry reduction in TiO activates additional three-phonon scattering channels~\cite{jin2024strain}. In layered MoS$_2$, modifications through defects and stacking variations significantly alter phonon--phonon, defect, and boundary scattering mechanisms, thereby tuning thermal conductivity~\cite{cammarata2019phonon}. Variations in layer number also influence recombination pathways through symmetry evolution from $P_3$ to $P_6$ structures~\cite{cammarata2019phonon}. More recently, topological phonons have emerged as a promising route for manipulating thermal transport. Analogous to topological electronic states, phonon spectra can host symmetry-protected Dirac, Weyl, nodal-line, and nodal-surface excitations~\cite{liu2019categories,zhang2023weyl,liu2022ubiquitous,jin2018recipe}. \textcolor{black}{Recent studies have systematically investigated the symmetry conditions, material realizations, and experimental signatures of Dirac- and Weyl-like phonons in both two- and three-dimensional crystalline systems\cite{liu2020symmetry,liu2025weyl,feng2022dirac,gong2022dirac}.} These topological features strongly influence phonon scattering and heat conduction. For instance, triple-point topological metals exhibit suppressed $\kappa_l$ arising from acoustic--optical phonon branch crossings~\cite{singh2018topological}, while rock-salt Sn$X$ compounds have been identified as type-II Weyl phononic systems~\cite{sihi2022evidence}.
\textcolor{black}{The reduction in thermal conductivity in materials caused by topological phonon states is also closely related to phonon softening induced by topological band crossings. Similar strategies for lowering thermal conductivity have also been employed in many thermoelectric materials such as half-Heusler compounds \cite{han2023strong,wang2025anomalous}, Mg$_3$Sb$_2$, and SnSe \cite{liu2026tellurium} }

In addition to conventional crystal structures, unconventional lattice geometries such as kagome, Lieb, Kekul\'e, square, and hyperhoneycomb lattices provide fertile platforms for realizing topological phonon states~\cite{wang2024topological,bhattacharya2019flat,xiong2022topological,mullen2015line,mu2022kekule}. Their geometric frustration and nonsymmorphic symmetries generate protected band crossings, flat modes, and unusual collective excitations. Among them, the square--octagonal (\textit{so}) lattice, composed of alternating square and octagonal units, is particularly attractive. Previous studies predicted Dirac fermions with high carrier mobility in \textit{so}-MoS$_2$~\cite{li2014gapless}. Although research has mainly focused on electronic topology, the distinctive symmetry and connectivity of the square--octagonal lattice are also expected to strongly modify phonon dispersions, potentially producing nodal lines, Dirac crossings, and tunable phonon gaps. Moreover, its porous structural framework may significantly influence phonon propagation by altering phonon velocities and scattering processes~\cite{wang2021systematic}.

Hexagonal MoS$_2$ exhibits comparatively large $\kappa_l$ due to strong covalent bonding and long phonon mean free paths, which limit its thermoelectric efficiency. This motivates the exploration of alternative lattice geometries capable of enhancing phonon scattering while preserving favorable electronic properties. Despite increasing interest in square--octagonal materials, the connection between their phonon topology and thermal transport properties remains largely unexplored. In this work, we investigate two chalcogenide monolayers in the square--octagonal geometry using first-principles calculations. Our results reveal significantly reduced lattice thermal conductivities of 18.67~W/mK for \textit{so}-MoS$_2$ and 4.00~W/mK for \textit{so}-SnS compared with their conventional hexagonal counterparts, indicating strong potential for related applications.

An important characteristic of the square--octagonal phonon spectra is the emergence of symmetry-protected nodal lines and Dirac points, which influence heat transport in different ways. Dirac-type crossings enhance the phonon group velocity ($v_g$) and can promote heat conduction, whereas nodal-line features suppress $v_g$ and enlarge the scattering phase space, thereby reducing phonon lifetimes and lowering $\kappa_l$. Since $\kappa_l$ is determined by the combined effects of phonon velocity, lifetime, frequency, and scattering strength, understanding these competing mechanisms is essential. Motivated by these considerations, we perform a systematic investigation of both harmonic and anharmonic phonon transport in two square--octagonal chalcogenide monolayers with \textit{P4} symmetry and compare their behavior with that of the conventional hexagonal phases.

\section{Computational Details}
First-principles calculations are carried out within the framework of density functional theory (DFT) using the projector-augmented wave (PAW) formalism~\cite{kresse1999ultrasoft}, as implemented in the \textit{Vienna Ab-Initio Simulation Package} (VASP)~\cite{kresse1996efficiency,kresse1996efficient}. Exchange--correlation effects are treated using the Perdew--Burke--Ernzerhof (PBE) generalized gradient approximation (GGA). A plane-wave cutoff energy of 500~eV together with a $9 \times 9 \times 1$ Monkhorst--Pack $k$-point grid is employed throughout the calculations. To eliminate spurious interactions between periodic replicas, a vacuum spacing of 20~\AA\ is introduced along the out-of-plane direction.

The electronic self-consistent calculations are converged to an energy tolerance of $10^{-8}$~eV, while atomic positions are fully relaxed until the residual Hellmann--Feynman forces become smaller than $10^{-4}$~eV/\AA. Long-range dispersion interactions are accounted for using the DFT-D3 correction scheme proposed by Grimme~\cite{grimme2010consistent}.

The harmonic interatomic force constants (IFCs) are obtained from density functional perturbation theory (DFPT) calculations together with the Born--Huang sum rules~\cite{born1996dynamical}. This treatment guarantees the correct quadratic behavior of the flexural acoustic phonon branch near the $\Gamma$ point in two-dimensional systems~\cite{eriksson2019hiphive}, which is essential for reliable predictions of $\kappa_l$~\cite{carrete2016physically}. Both harmonic and third-order anharmonic IFCs are computed using a $3 \times 3 \times 1$ supercell with atomic interaction cutoffs extending up to 0.55~nm and a corresponding $2 \times 2 \times 1$ $k$-point mesh.

The lattice thermal conductivity ($\kappa_l$) is evaluated by solving the linearized phonon Boltzmann transport equation (BTE) iteratively beyond the relaxation time approximation. Three-phonon scattering processes are explicitly included through the second- and third-order IFCs~\cite{li2014shengbte}. To ensure fully converged thermal transport properties, a dense $80 \times 80 \times 1$ $q$-point mesh is adopted for the Brillouin-zone sampling.

\section{Results and Discussions}
\subsection{Crystal Structure}

\begin{figure}[h]
\centering
\includegraphics[scale=0.054]{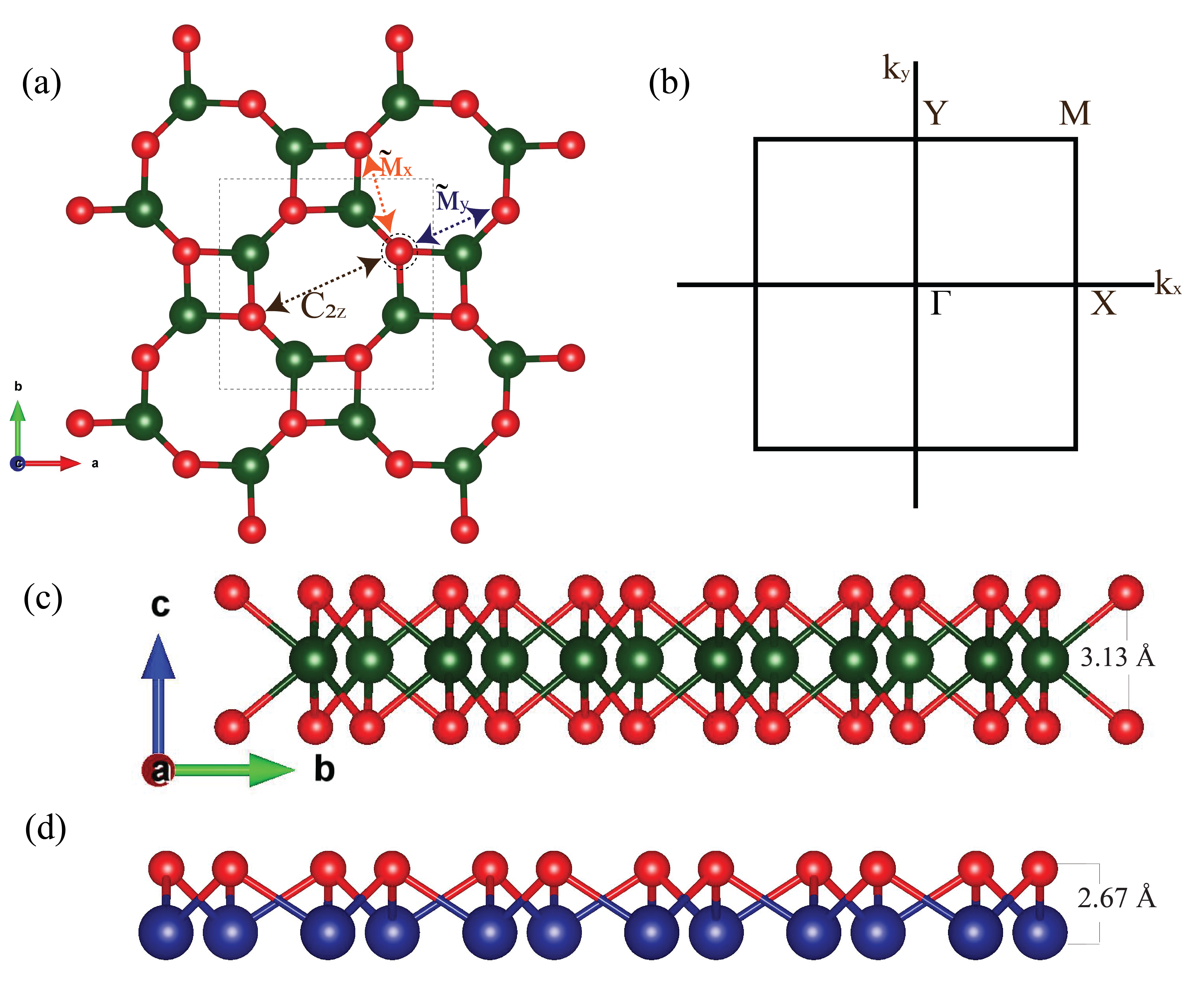} 
\caption{Crystal structures and corresponding Brillouin zones of the investigated monolayers. 
(a) Top view (along the $z$ direction) of \textit{so}-MoS$_2$ and \textit{so}-SnS monolayers. 
(b) Brillouin zone showing the high-symmetry points. 
Side views (along the $x$ direction) of (c) \textit{so}-MoS$_2$ and (d) \textit{so}-SnS. 
Mo, Sn, and S atoms are shown as green, blue, and red spheres, respectively. The dashed black square in panel (a) denotes the primitive unit cell. 
The important symmetry operations illustrated in panel (a) include the twofold rotation about the $z$ axis, 
$C_{2z} : (x,y) \rightarrow (-x,-y)$, 
the glide mirror symmetry along $x$, 
$\tilde{M}_x : (x,y) \rightarrow (-x+\frac{1}{2},\, y+\frac{1}{2})$, 
and the glide mirror symmetry along $y$, 
$\tilde{M}_y : (x,y) \rightarrow (x+\frac{1}{2},\, -y+\frac{1}{2})$. 
The two monolayers differ only by the presence of the mirror symmetry 
$M_z : (x,y,z) \rightarrow (x,y,-z)$, 
which exists in \textit{so}-MoS$_2$ but is absent in \textit{so}-SnS.
}
\label{crystalstructure}
\end{figure}

The studied monolayers crystallize in square--octagonal (\textit{so}) frameworks composed of alternating square and octagonal rings. The \textit{so}-MoS$_2$ monolayer belongs to the tetragonal P\textit{4mbm} space group, whereas \textit{so}-SnS adopts the non-centrosymmetric P$4bm$ structure. In \textit{so}-MoS$_2$, the primitive unit cell contains four Mo and eight S atoms occupying the 4h and 8k Wyckoff positions, respectively. Each Mo atom is coordinated by surrounding S atoms, while every S atom bonds with three neighboring Mo atoms (see Fig.~\ref{crystalstructure} and Fig.~\ref{polyhedra_MoS2} in the supplementary material (SM)~\cite{supp}). Similar to hexagonal MoS$_2$, the structure exhibits a trilayer arrangement with the Mo layer sandwiched between two S layers. The optimized lattice parameter and layer thickness are calculated to be 6.36~\AA\ and 3.12~\AA, respectively. In contrast to the uniform Mo--S bond length of 2.42~\AA\ in the hexagonal phase, the square--octagonal geometry produces two distinct bond lengths of 2.46~\AA\ and 2.39~\AA\ associated with the square and octagonal motifs\textcolor{black}{(see Table\ref{hse_SnX}) in SM\cite{supp})}

The \textit{so}-MoS$_2$ lattice belongs to the centrosymmetric D$_{4h}$ (4/mmm) point group and possesses C$_4$ and C$_2$ rotational symmetries along the $z$ direction together with mirror and diagonal reflection planes. In addition, glide symmetries generate a non-symmorphic crystal structure, leading to Brillouin-zone folding and symmetry-enforced phonon degeneracies~\cite{juneja2021quasiparticle,aiswarya2025symmetry}.

\begin{figure}[h]
\centering
\includegraphics[scale=0.03]{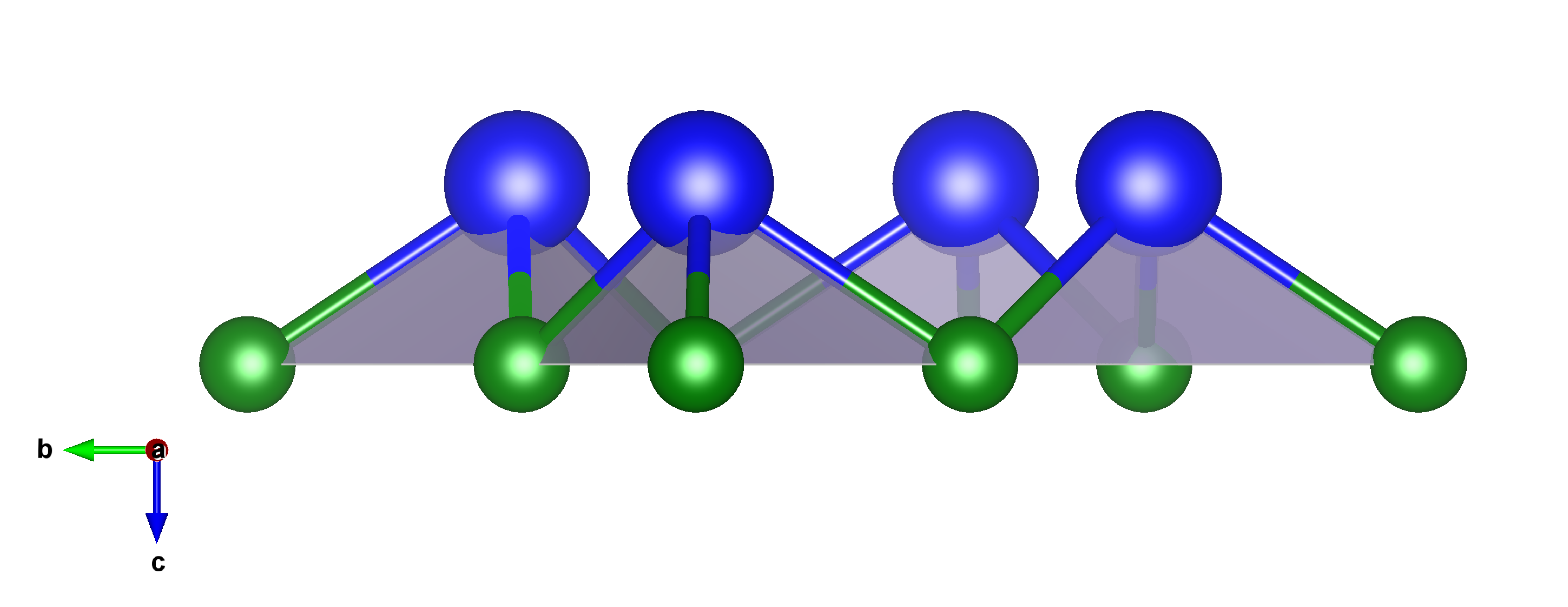} 
\caption{The trigonal pyramidal configuration of \textit{so}-SnS monolayer.}
\label{polyhedra-SnX}
\end{figure}

Unlike \textit{so}-MoS$_2$, the \textit{so}-SnS monolayer lacks inversion symmetry and crystallizes in the P$4bm$ space group corresponding to the C$_{4v}$ (4mm) point group. The structure preserves C$_4$ and C$_2$ rotational operations as well as mirror and glide symmetries, which support symmetry-protected phonon band crossings, including nodal-line and Weyl-like degeneracies. Its primitive cell contains four Sn and four S atoms with optimized lattice constant and thickness values of 7.41~\AA\ and 2.67~\AA, respectively. Structurally, Sn atoms form a distorted square-planar arrangement within the $ab$ plane, while S atoms are displaced along the $c$ direction, giving rise to trigonal pyramidal coordination around Sn atoms (see Fig.~\ref{crystalstructure} and Fig.~\ref{polyhedra-SnX} in the SM~\cite{supp}). The monolayer further exhibits a bilayer-like stacking character with Sn--S bond lengths of 2.53~\AA\ in the octagonal units and 2.67~\AA\ in the square units, both smaller than those reported for hexagonal and orthorhombic SnS phases~\cite{nair2023metavalent}.

Owing to the tetragonal symmetry of the square--octagonal lattice, the in-plane elastic behavior is primarily determined by the elastic constants $C_{11}$ and $C_{12}$, with $C_{66}$ satisfying the relation $C_{66} = (C_{11} - C_{12})/2$. The calculated elastic constants satisfy the Born mechanical stability criteria~\cite{wang2022high} (see Table~\ref{elastic_so-MX} in the SM~\cite{supp}), confirming the structural stability and mechanical robustness of both monolayers.

\subsection{Symmetry Analysis and Protected Band Crossings }

\begin{figure}[t]
\centering
\includegraphics[scale=0.09]{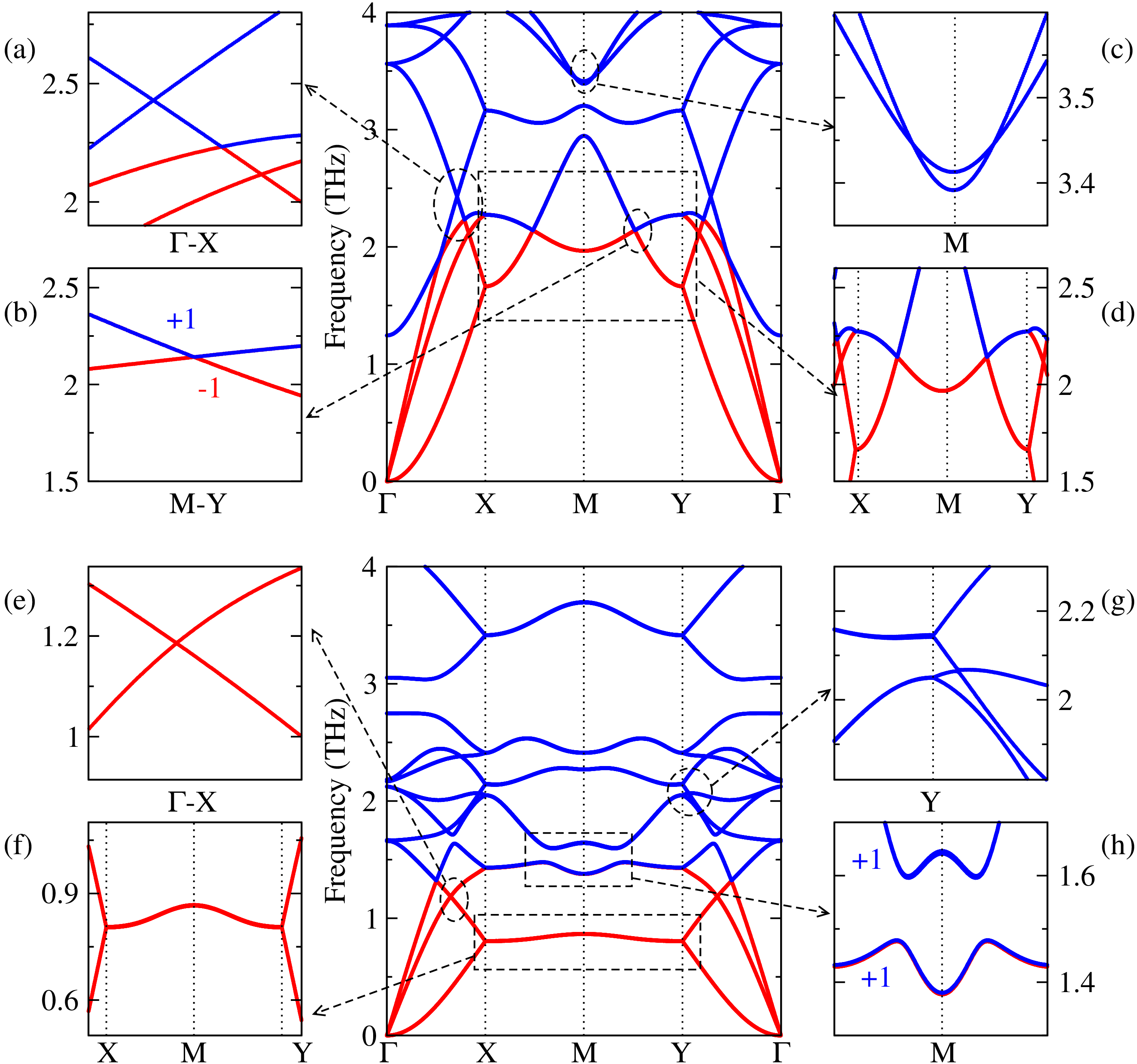} 
\caption{Phonon band structures and symmetry-protected band crossings in (a--d) \textit{so}-MoS$_2$ and (e--h) \textit{so}-SnS monolayers. Acoustic and optical phonon branches are shown by red and blue curves, respectively. Enlarged views of representative phonon crossings are presented in panels (a--d) for \textit{so}-MoS$_2$ and panels (e--h) for \textit{so}-SnS. (a) Magnified view of a twofold band crossing along the $\Gamma$--$X$ direction.  (b) Enlarged view of a fourfold band crossing along the $M$--$Y$ direction. The labels $+1$ and $-1$ denote the eigenvalues of the glide mirror operator $\tilde{M}_x$. 
(c,d) Detailed views of the highlighted regions in the phonon dispersion of \textit{so}-MoS$_2$.  (e--h) Enlarged views of representative phonon crossings and avoided crossings in the phonon spectrum of \textit{so}-SnS. }
\label{topologicalcrossing_MoS2}
\end{figure}

\subsubsection{Symmetry Analysis}
In this section, we analyze the microscopic origin of the point and line degeneracies appearing in the phonon spectra of the square--octagonal lattices. Such band crossings may originate either from crystalline symmetry or from topological constraints imposed by symmetry operations in momentum space. We therefore first examine the symmetry properties of \textit{so}-MoS$_2$ and \textit{so}-SnS and their consequences for the phonon band structure.
Both monolayers crystallize in non-symmorphic space groups containing fractional lattice translations along the $x$ and $y$ directions together with the $D_{4h}$ and $D_{4v}$ point groups, respectively. The principal distinction between the two systems is the presence or absence of the mirror operation $M_z : (x,y,z)\rightarrow(x,y,-z)$ (see Fig.~\ref{crystalstructure}(c,d)). Since the systems are strictly two-dimensional, the symmetry analysis can be restricted to the $k_x$--$k_y$ plane. Within this plane, the operation $M_z$ leaves the in-plane momentum unchanged, implying that the symmetry-protected degeneracies discussed below have the same origin in both materials. The coexistence of point-group symmetries and fractional translations gives rise to characteristic non-symmorphic band degeneracies. The relevant symmetry operations are summarized below.

\begin{enumerate}
\item[(i)] The glide mirror symmetry
\[
\tilde{M}_x = \{ M_x \,|\, \mathbf{t} \},
\]
which combines mirror reflection with a fractional translation. Its action in real and momentum space is
\begin{eqnarray}
\tilde{M}_x &:& (x,y) \rightarrow \left(-x + \frac{1}{2},\, y + \frac{1}{2}\right), \nonumber\\
\tilde{M}_x &:& (k_x,k_y) \rightarrow (-k_x,k_y). \nonumber
\end{eqnarray}

\item[(ii)] The glide mirror symmetry $\tilde{M}_y$,
\begin{eqnarray}
\tilde{M}_y &:& (x,y) \rightarrow \left(x + \frac{1}{2},\, -y + \frac{1}{2}\right), \nonumber\\
\tilde{M}_y &:& (k_x,k_y) \rightarrow (k_x,-k_y). \nonumber
\end{eqnarray}

\item[(iii)] A twofold rotational symmetry about the $z$ axis,
\[
C_{2z} : (x,y) \rightarrow (-x,-y),
\qquad
C_{2z} : (k_x,k_y) \rightarrow (-k_x,-k_y).
\]

\item[(iv)] Time-reversal symmetry ($T$), which is preserved because the systems are nonmagnetic. In momentum space,
\[
T : (k_x,k_y) \rightarrow (-k_x,-k_y).
\]
For spinless phonons, $T^2 = +1$, and in a suitable basis, one may write $T=\mathcal{K}$, where $\mathcal{K}$ denotes complex conjugation.
\end{enumerate}

These symmetry operations constrain the Bloch Hamiltonian and enforce degeneracies at selected high-symmetry points and lines within the Brillouin zone. In particular, the glide symmetries $\tilde{M}_x$ and $\tilde{M}_y$ produce momentum-dependent eigenvalues whose algebra, combined with $C_{2z}$ and $T$, leads to symmetry-protected Dirac points and nodal-line crossings in both electronic and phononic band structures.

We next show how these symmetries generate topological nodal features in the proposed systems. Along the high-symmetry path $X-M-Y$, all phonon branches become doubly degenerate, forming symmetry-enforced twofold nodal lines, as illustrated in Fig.~\ref{topologicalcrossing_MoS2}. In addition, the intersection of two such nodal lines may produce a fourfold band crossing identified as a Dirac node. To clarify the origin of these degeneracies, consider a Bloch state $\ket{\psi}=e^{i\mathbf{k}\cdot\mathbf{r}}u_{\mathbf{k}}(\mathbf{r})$ and apply the glide operation $\tilde{M}_x$ twice. Since $\tilde{M}_x$ contains a fractional translation along the $y$ direction, one obtains
\[
\tilde{M}_x^2 = e^{ik_y}.
\]
Along the $M-Y$ direction, where $k_y=\pi$, this relation becomes $\tilde{M}_x^2=-1$. Consequently, the anti-unitary operator $\tilde{M}_xT$ satisfies
\[
(\tilde{M}_xT)^2=-1,
\]
which enforces a local Kramers-like degeneracy even in the absence of spin--orbit coupling. Therefore, every band along the $M-Y$ line is at least doubly degenerate, giving rise to an essential nodal line. The corresponding orthogonal states are $\ket{\psi}$ and $(\tilde{M}_xT)\ket{\psi}$.

When two nodal lines intersect along a high-symmetry direction, a fourfold Dirac crossing can emerge. If the intersecting bands belong to different $\tilde{M}_x$ eigenvalue sectors ($+1$ and $-1$), hybridization between them is symmetry forbidden. Their crossing, therefore, remains protected, producing a fourfold Dirac node along the $M-Y$ direction, as shown in Fig.~\ref{topologicalcrossing_MoS2}(b) within the frequency range of 1.5--2.5~THz. In this case, one nodal line formed by the pair $\ket{\psi}$ and $(\tilde{M}_xT)\ket{\psi}$ belongs to the $\tilde{M}_x=+1$ sector, while the second belongs to the $\tilde{M}_x=-1$ sector. It should be noted that these fourfold Dirac nodes are accidental rather than symmetry-enforced. Depending on the symmetry compatibility and band ordering, both protected crossings and avoided crossings may occur along the $M-Y$ path, as illustrated in Fig.~\ref{topologicalcrossing_MoS2}(d,h).

Along the $k_y=0$ direction corresponding to the $\Gamma-X$ path, the Kramers-like condition is absent, and the bands are generally nondegenerate (see Fig.~\ref{topologicalcrossing_MoS2}). Nevertheless, symmetry-protected twofold crossings, referred to here as Weyl nodes, may still appear. One possible protection mechanism originates from the mirror symmetry $M_x$. Two nondegenerate branches carrying opposite $M_x$ eigenvalues ($\pm1$) cannot hybridize and therefore form a protected crossing upon intersection. A similar symmetry analysis can be applied along the $M-X$ direction by considering the combined effects of $T$, $M_y$, and the fractional lattice translations, which likewise generate essential nodal lines and accidental Dirac crossings.

Another important symmetry in these systems is the combined anti-unitary operation $C_{2z}T$. Unlike ordinary time-reversal symmetry, this operation leaves the momentum invariant and is therefore local in momentum space. On a suitable basis, one may represent $C_{2z}T=\mathcal{K}$, which constrains the Hamiltonian to be purely real. Under this symmetry, a generic two-band Hamiltonian in two dimensions can be written as
\[
H(\mathbf{k}) = f(\mathbf{k})\sigma_x + g(\mathbf{k})\sigma_z,
\]
where $f(\mathbf{k})$ and $g(\mathbf{k})$ are real functions of $(k_x,k_y)$ and $\sigma_x$, $\sigma_z$ are Pauli matrices. A Weyl point emerges when both $f=0$ and $g=0$, provided the $C_{2z}T$ symmetry remains intact. Additional crystalline symmetries such as mirror or rotational operations may further pin these crossings to specific high-symmetry points or lines in the Brillouin zone, analogous to the symmetry-enforced Dirac cones in graphene~\cite{RevModPhys.81.109}. The $C_{2z}T$ symmetry can also stabilize both linear and quadratic band touchings characterized by topological winding numbers of $\pm1$ and $\pm2$, respectively~\cite{PhysRevLett.133.196603,3pnm-76hh}. Although the total winding number over the Brillouin zone must vanish in a strict two-band model, isolated subspaces in multiband systems may still carry finite winding numbers~\cite{PhysRevX.9.021013}. Similar symmetry analysis is presented in other material candidates elsewhere \cite{Slager1,Slager2,Slager3}

\subsection{Phonon Dispersion}

\begin{figure}[t]
\centering
\includegraphics[scale=0.28]{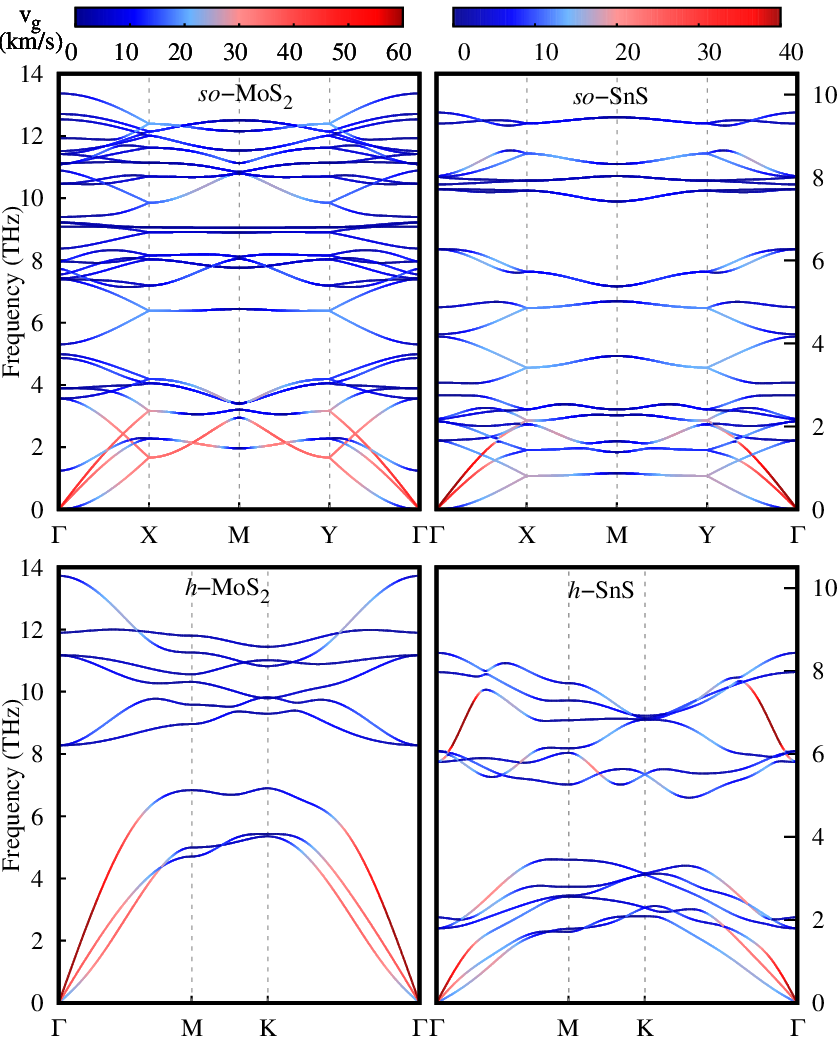} 
\caption{\textcolor{black}{Calculated phonon band dispersions of MoS$_2$ and SnS monolayers in their $h$- and $so$-phases, along the high symmetry directions of their respective Brillouin zones. The phonon branches are color-mapped according to the magnitude of the phonon group velocity (v$_g$), as indicated by the color scale on the top. } }
\label{phononbanddispersion}
\end{figure}

Figure \ref{phononbanddispersion} present the phonon dispersions of MoS$_2$ and SnS monolayers in their \textcolor{black}{hexagonal ($h$) and square octagonal ($so$) phases} along the high-symmetry directions of their respective Brillouin zones. The phonon branches are color-coded according to the magnitude of the phonon group velocity, as indicated by the color scale on the top. The absence of imaginary frequencies throughout the Brillouin zone confirms the dynamical stability of both structures. The primitive unit cells of \textit{so}-MoS$_2$ and \textit{so}-SnS contain 12 and 8 atoms, respectively, corresponding to 36 phonon branches (3 acoustic and 33 optical) for \textit{so}-MoS$_2$ and 24 branches (3 acoustic and 21 optical) for \textit{so}-SnS.

Both systems exhibit strong hybridization between acoustic and low-energy optical phonon modes. Such coupling opens additional phonon scattering pathways, decreases phonon lifetimes, and consequently suppresses thermal transport. In \textit{so}-MoS$_2$, the non-symmorphic crystal symmetry generates doubly degenerate phonon branches along the $X$--$M$--$Y$ path, forming twofold nodal lines. These degeneracies appear in both the acoustic and low-frequency optical regions and therefore play an important role in phonon transport. A fourfold degenerate point is observed near 2.13~THz, originating from the crossing of two doubly degenerate branches, as illustrated in Fig.~\ref{topologicalcrossing_MoS2}(b). Additional symmetry-protected twofold nodal crossings are found around 2.3~THz along the $\Gamma$--$X$ and $M$--$\Gamma$ directions, stabilized by mirror symmetry.

Compared with conventional hexagonal MoS$_2$, which possesses relatively high $\kappa_l$ due to the presence of a phonon forbidden gap and weak acoustic--optical coupling, the square--octagonal phase displays several characteristics favorable for suppressing heat transport. Similar symmetry-protected degeneracies and enhanced acoustic--optical interactions are also present in \textit{so}-SnS. In particular, the closure of phonon gaps together with the softening of acoustic branches substantially lowers phonon lifetimes. Along the $X$--$M$ direction, several weakly dispersive flat branches are observed, leading to reduced phonon group velocities defined by $v_g=\nabla_k\omega$. The nodal-line-like features further flatten portions of the dispersion spectrum, whereas linearly dispersive branches and Dirac-cone-like crossings along the $\Gamma$--$X$ and $M$--$\Gamma$ directions locally enhance $v_g$. The competition between these features strongly influences the overall $\kappa_l$.

\textcolor{black}{The lattice stiffness can be characterized by the Debye temperature ($\theta_D$) which is closely related to the sound velocity and acoustic phonon spectrum ~\cite{denault2015average,tohei2006debye}. The Debye temperature is defined as
\begin{equation}
\theta_D = \frac{\hbar \omega_D}{k_B},
\end{equation}
where $\hbar$ is the reduced Planck constant, $\omega_D$ the maximum acoustic phonon frequency (or Debye frequency), and $k_B$ is the Boltzmann constant. The calculated $\omega_D$ and $\theta_D$ for both the monolayers in their $h$ and $so$-phases are shown in Table~\ref{thetaD}. It is quite evident from the  $\theta_D$ values that $so$-lattice is significantly softer than the hexagonal one ($\theta_D$ for \textit{so}-MoS$_2$ and \textit{so}-SnS are approximately 53\% and 40\% smaller than those of the corresponding hexagonal phases), leading to lower $\kappa_l$. }

\begin{table}[htbp]
\centering
\renewcommand{\arraystretch}{1.5}
\caption{Calculated Debye frequency ($\omega_D$) and Debye temperature
($\theta_D$) of MoS$_2$ and SnS monolayers in hexagonal ($h$) and
\textit{so} phases.}
\label{thetaD}

\begin{tabular}{|c|c|c|c|c|}
\hline
\textbf{Monolayer} & $h$-MoS$_2$ & $h$-SnS &
\textit{so}-MoS$_2$ & \textit{so}-SnS \\
\hline
$\omega_D$ (THz) & 6.80 & 2.50 & 2.9 & 1.50 \\
$\theta_D$ (K) & 326 & 120 & 139 & 72 \\
\hline
\end{tabular}
\end{table}

\subsection{Lattice Thermal Transport}

\begin{figure}[t]
\centering
\includegraphics[scale=0.12]{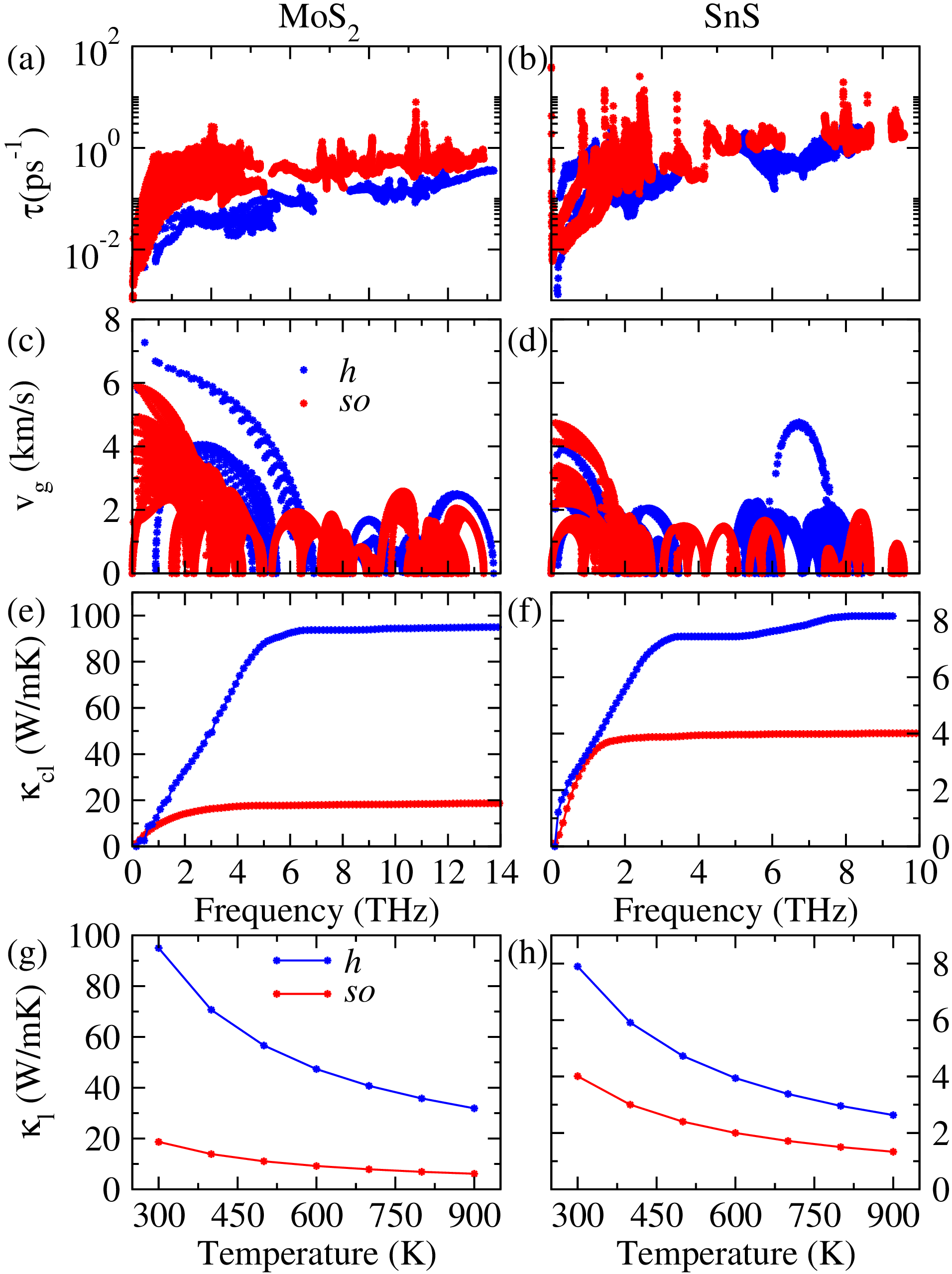}
\caption{For the $h$- and $so$-phases of MoS$_2$ and SnS monolayers, calculated (a,b) three-phonon anharmonic scattering rates ($\tau$), (c,d) phonon group velocity ($v_g$), (e,f) cumulative lattice thermal conductivity ($\kappa_{cl}$) as a function of phonon frequency, and (g,h) lattice thermal conductivity ($\kappa_l$) as a function of temperature.}
\label{kappa_l}
\end{figure}

The lattice thermal transport in square--octagonal ($so$) chalcogenides is intrinsically governed by the underlying lattice symmetry, which dictates both the harmonic phonon dispersion and the anharmonic phonon scattering processes. Compared with the conventional hexagonal ($h$) phase, the $so$ lattice introduces multiple symmetry-protected and accidental phonon band crossings that significantly modify the low-energy phonon spectrum. As discussed in the previous section, these topological features reshape the phonon dispersion through band flattening, enhanced acoustic--optical hybridization, and phonon softening, collectively reducing the efficiency of heat transport.
Figures~\ref{kappa_l}(a,b) compare the frequency-dependent three-phonon scattering rates ($\tau$) of the $h$- and $so$-phases of MoS$_2$ and SnS monolayers. \textcolor{black}{A pronounced enhancement of anharmonic scattering is observed in both $so$ structures. This enhancement originates primarily from the increased number of acoustic--optical band crossings, which promote stronger mode hybridization, increase the number of allowed three-phonon processes, and substantially enlarge the anharmonic scattering phase space~\cite{li2016influence}.} Consequently, the maximum scattering rate of $so$-MoS$_2$ reaches nearly 3~ps$^{-1}$, more than an order of magnitude larger than that of its $h$ counterpart. A similar trend is found in $so$-SnS, where the strong acoustic--optical coupling below 3~THz (particularly below 2~THz) gives rise to scattering rates of approximately 39~ps$^{-1}$ (100~ps$^{-1}$), exceeding those of the corresponding $h$ phase by more than one order of magnitude.

The influence of phonon topology extends beyond anharmonic scattering by modifying the phonon group velocity ($v_g$), as illustrated in Figs.~\ref{kappa_l}(c,d). Dirac-like crossings retain relatively large $v_g$ because of their linear dispersions, whereas nodal-line segments become weakly dispersive and locally suppress the group velocity. It is important to note that the nodal lines themselves do not directly reduce $v_g$; rather, the accompanying band flattening decreases the dispersion slope ($v_g=\nabla_k\omega$), producing an enhanced phonon density of states over a narrow frequency window (see Fig. S4). The resulting increase in the available three-phonon scattering channels further strengthens phonon--phonon interactions and shortens phonon lifetimes. This reduction in $v_g$ is particularly pronounced in the low-frequency region of $so$-MoS$_2$, whereas for $so$-SnS it is more evident at relatively higher phonon frequencies. Therefore, the remarkably low lattice thermal conductivity of the $so$ phases originates from the combined effects of reduced phonon group velocity and enhanced anharmonic scattering induced by symmetry-protected phonon topology.

\begin{figure}[h]
\centering
\includegraphics[width=0.9\linewidth]{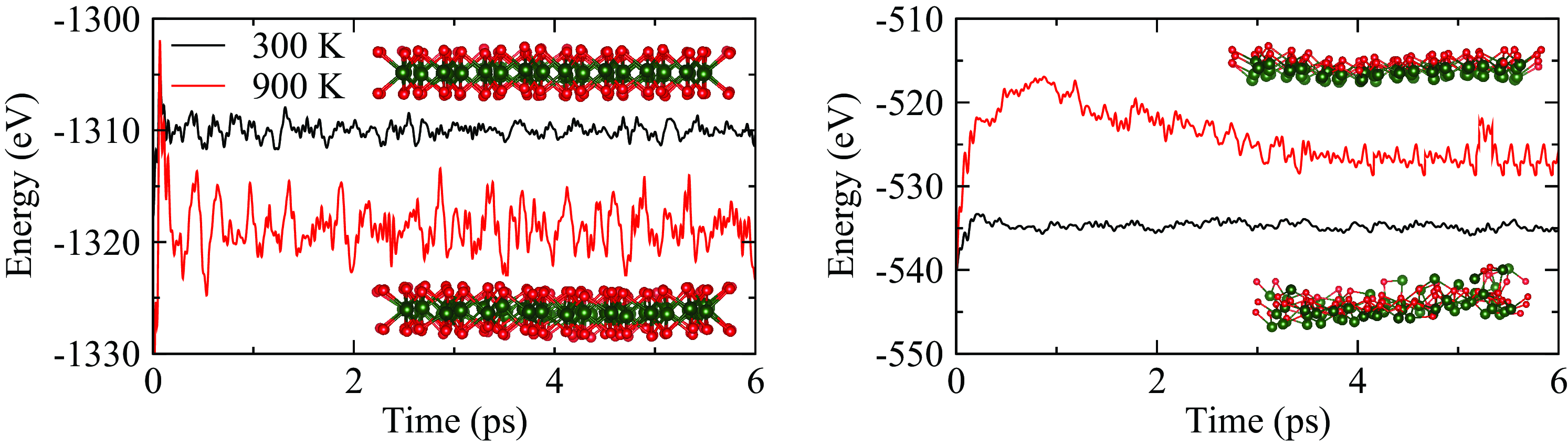} 
\caption{Calculated thermodynamic stability of \textit{so} monolayers using AIMD. The top panel shows the optimized crystal structure after equilibration at 300 K, and the bottom panel displays the configuration obtained at 900 K.}
\label{AIMD}
\end{figure}

The combined influence of suppressed phonon group velocity and enhanced anharmonic scattering is directly reflected in the lattice thermal conductivity shown in Figs.~\ref{kappa_l}(g,h). For all four systems, $\kappa_l$ decreases monotonically with increasing temperature owing to the progressive strengthening of Umklapp phonon scattering. Although $so$-SnS becomes thermodynamically unstable at elevated temperatures, its calculated thermal conductivity is presented for comparison (see Fig.~\ref{AIMD}). \textcolor{black}{ The room-temperature lattice thermal conductivities of the $h$-MoS$_2$ and $h$-SnS monolayers are calculated to be 98.00 and 8.00~W/mK, respectively, in excellent agreement with previous theoretical reports~\cite{nair2023metavalent,gandi2016thermal,nair2025ultra}, confirming the reliability of our computational approach. In comparison, the room-temperature $\kappa_l$ of $so$-MoS$_2$ is reduced to 18.67~W/mK, nearly eight times smaller than that of the hexagonal phase~\cite{nair2025ultra,gandi2016thermal}, while $so$-SnS exhibits a $\kappa_l$ of only 4.00~W/mK, corresponding to an approximately twofold reduction compared with $h$-SnS~\cite{nair2023metavalent}.}

\begin{figure}[h]
\centering
\includegraphics[width=0.5\linewidth]{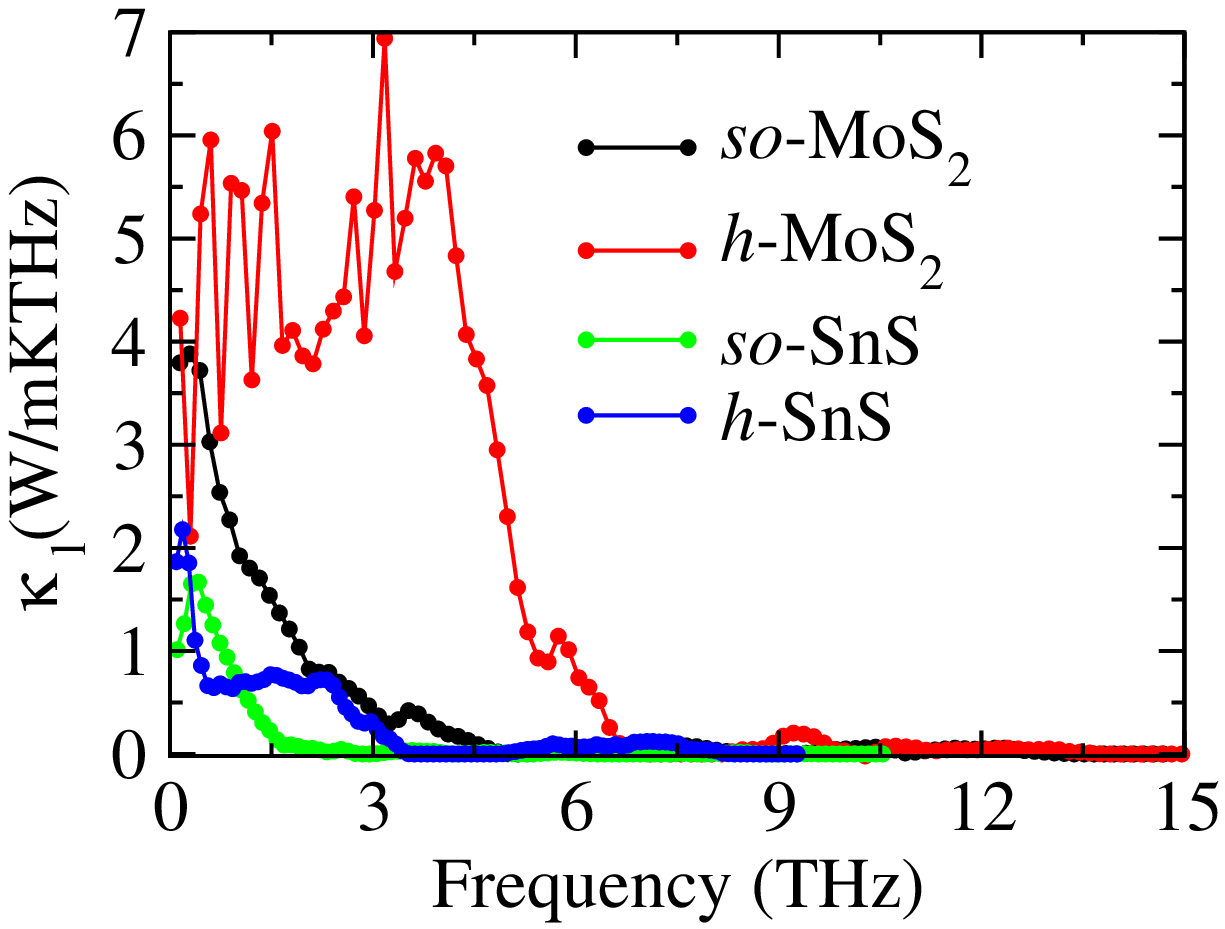} 
\caption{Calculated spectral decomposition of $\kappa_l$ of MoS$_2$ and SnS monolayers in their hexagonal ($h$) and $so$ lattices.}
\label{spectral_so}
\end{figure}

\textcolor{black}{The remarkable suppression of $\kappa_l$ in the $so$ phases originates from the combined effects of symmetry-protected and accidental phonon band crossings, phonon softening, acoustic phonon bunching, and enhanced acoustic--optical mode coupling, all of which substantially increase anharmonic phonon scattering. Similar phonon-engineering strategies based on phonon softening and avoided phonon-band crossings have recently been employed to suppress lattice thermal conductivity in thermoelectric materials such as half-Heusler compounds, Mg$_3$Sb$_2$, and SnSe~\cite{liu2026tellurium,han2023strong,wang2025anomalous}.} Compared with the $h$ phase, the $so$ lattice exhibits a considerably larger number of low-energy phonon branches and a much stronger overlap between acoustic and optical modes, thereby promoting phonon scattering throughout the low-frequency region. In particular, the absence of the phonon band gap in $so$-MoS$_2$ further strengthens acoustic--optical scattering, while its relatively low Debye temperature (139~K) and approximately 35\% contribution from flexural acoustic phonons suppress heat transport even further. Consequently, spectral decomposition of $\kappa_l$ (Fig.~\ref{spectral_so}) shows that phonons above 4.5~THz contribute negligibly to heat conduction in $so$-MoS$_2$, whereas the corresponding $h$ phase retains appreciable contributions beyond 6~THz. In contrast, the larger phonon gap, higher Debye temperature ($\Theta_D\sim300$~K), and absence of low-energy topological phonon features enable significantly more efficient heat transport in $h$-MoS$_2$.

\begin{figure}[h]
\centering
\includegraphics[width=0.5\linewidth]{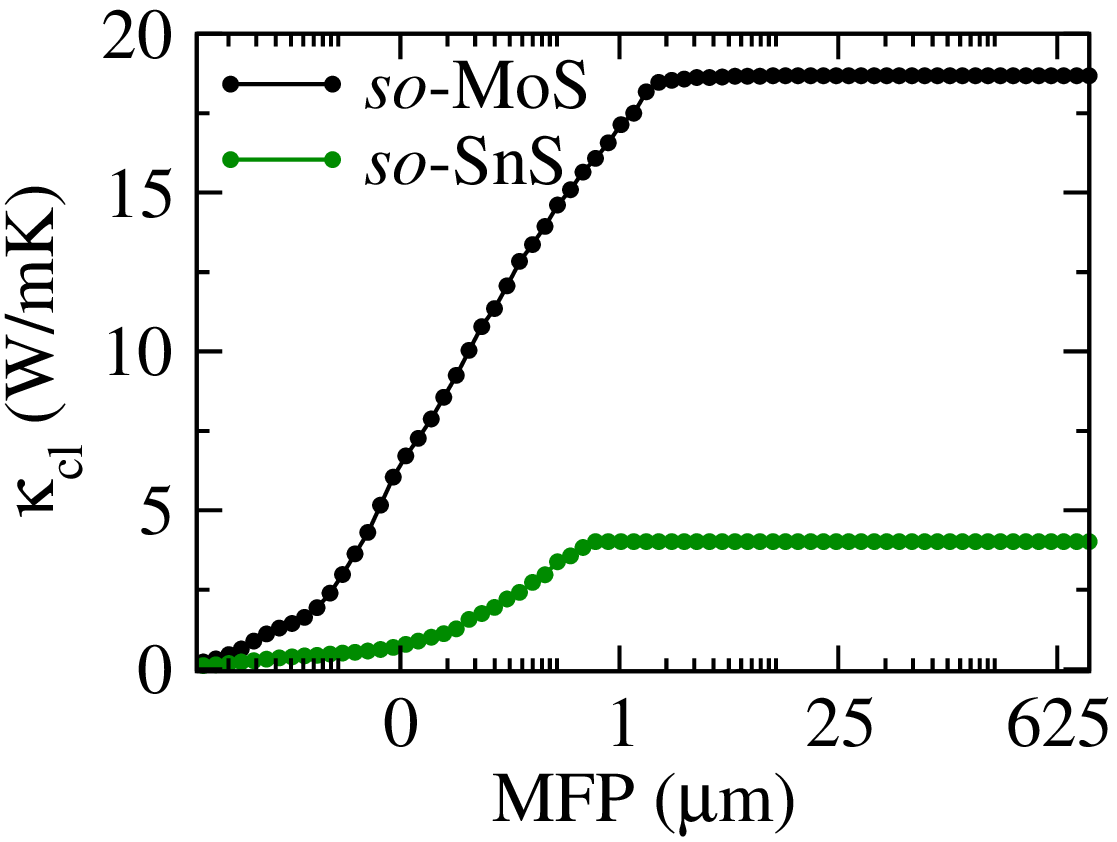} 
\caption{\textcolor{black}{Calculated cumulative lattice thermal conductivity ($\kappa_{cl}$) of $so$-MoS$_2$ and so-SnS monolayers vs. mean free path (MFP).}}
\label{revision_kcl}
\end{figure}

A similar microscopic picture emerges for $so$-SnS. Its reduced thermal conductivity originates from the lower Debye temperature, multiple phonon band crossings below 1.5~THz, and the strong suppression of high-frequency phonon transport arising from enhanced anharmonic scattering and reduced phonon group velocities. \textcolor{black}{As illustrated by the cumulative lattice thermal conductivity ($\kappa_{cl}$) in Figs.~\ref{kappa_l}(e,f) and \ref{spectral_so}, heat conduction in $so$-SnS is almost entirely dominated by phonons below 1.5~THz, whereas the $h$ phase receives substantial contributions from phonons above 3~THz~\cite{nair2023metavalent}.} Figure~\ref{revision_kcl} further demonstrates that nearly 90\% of the total lattice thermal conductivity is contributed by phonons with mean free paths shorter than 0.7~$\mu$m in $so$-MoS$_2$ and 0.5~$\mu$m in $so$-SnS. This strong localization of heat-carrying phonons also explains the pronounced size dependence of $\kappa_l$. At characteristic sample lengths of 2~$\mu$m for $so$-MoS$_2$ and 0.76~$\mu$m for $so$-SnS, the calculated thermal conductivity approaches the bulk limit. However, reducing the sample dimension to 100~nm decreases $\kappa_l$ to approximately 9.5~W/mK and 1.4~W/mK, respectively, corresponding to more than a twofold reduction. This additional suppression originates from enhanced boundary scattering, which becomes increasingly effective once the sample dimensions approach the intrinsic phonon mean free paths.

\section{Conclusions}
Unconventional square--octagonal lattices offer an attractive platform for realizing symmetry-protected topological phonon states beyond those found in conventional crystalline materials. In this work, we demonstrated that square--octagonal chalcogenide monolayers host nontrivial phonon band topologies that play a central role in determining lattice thermal transport. Symmetry-protected band crossings, weakly dispersive phonon branches, and anisotropic phonon modes collectively reshape the phonon group velocities and scattering channels, leading to strongly suppressed lattice thermal conductivity.
Using \textit{so}-MoS$_2$ and \textit{so}-SnS monolayers as representative systems, we showed that thermal transport is fundamentally controlled by the interplay between glide-mirror and time-reversal symmetries, which enforce nodal-line degeneracies whose intersections generate Dirac nodes. Our analysis further reveals that nodal lines themselves do not directly suppress the phonon group velocity ($v_g$); instead, the associated band flattening locally reduces $v_g$ and modifies the phonon density of states. As a consequence, lattice heat transport is governed by the combined effects of phonon velocity and enhanced phonon scattering, extending beyond the conventional picture of thermal conduction.
In addition, the closure of the phonon forbidden gap in \textit{so}-MoS$_2$ and \textit{so}-SnS substantially enhances phonon--phonon scattering rates by nearly 30 and 15 times, respectively. This leads to remarkably low room-temperature lattice thermal conductivities of 18.67~W/mK for \textit{so}-MoS$_2$ and 4.00~W/mK for \textit{so}-SnS, corresponding to reductions of approximately 8-fold and 2-fold, respectively, compared with their hexagonal counterparts.

Unlike conventional approaches based on defects, interfaces, alloying, or nanostructuring, our results demonstrate an intrinsic mechanism for engineering thermal transport that originates directly from lattice geometry and topological phonon features. These findings establish topological phonon dispersion as an effective and previously underexplored design principle for phononic materials.
More broadly, the present study advances the fundamental understanding of heat transport in low-dimensional systems and provides useful guidelines for the design of next-generation thermoelectric and thermal management materials. Our work further suggests that incorporating topological concepts into phonon engineering may enable robust and controllable thermal functionalities in future nanoelectronic and quantum devices.

\section{Conflicts of Interest} 
The authors declare no conflict of interest.\\

\section{Data Availability} 
Data supporting the study are included within the article and
supplementary material (SM). Supplementary information provides further auxiliary details about the polyhedral structure of $so$-MoS$_2$, the trigonal pyramidal configuration of $so$-SnS monolayer, phonon DOS of $so$-MoS$_2$ and $so$-SnS, and their respective thermodynamic stability using molecular dynamics calculations, calculated lattice thermal conductivity of MoS$_2$ and SnS in their hexagonal and $so$-lattices, simulated elastic constants, bond length, bond stiffness, and HSE06 band gaps.  
  
\section{Acknowledgment} 
  We acknowledge the use of the high-performance computing resources and support provided by Almesbar at Khalifa University of Science and Technology. This research was funded by Khalifa University of Science and Technology through the RIG under Project ID: KU-INT-RIG-2023-01-8474000554. C.M. was supported by Samsung Science and Technology Foundation under project no. SSTF-BA2002-06, National Research Foundation of Korea (NRF) funded by the Korean government(MSIT),  grant no. RS-2021-NR060087 and RS-2025-00562579,  Global Research Development-Center (GRDC) Cooperative Hub Program through the NRF funded by the MSIT, grant no. RS-2023-00258359, Global-LAMP program of the NRF funded by the Ministry of Education, grant no. RS-2023-00301976.

\bibliography{Reference}
\clearpage

\onecolumngrid

\newpage
\begin{center}
{\Large \bfseries Supplementary Materials\par}

\vspace{0.5em}

{\Large \bfseries
Symmetry-Protected Phonon Topology and Low Lattice Thermal Conductivity in Square–Octagonal Chalcogenides\par}

\vspace{1.5em}

Surabhi Suresh Nair\\
{\itshape Department of Physics, Khalifa University, Abu Dhabi-127788, United Arab Emirates (UAE)}

\vspace{1em}

Chiranjit Mondal\\
{\itshape Department of Physics and Astronomy, Seoul National University, Seoul 08826, Korea\\
Center for Theoretical Physics (CTP), Seoul National University, Seoul 08826, Korea}

\vspace{1em}

Aftab Alam\textsuperscript{*}\\
{\itshape Department of Physics, Indian Institute of Technology, Bombay, Powai, Mumbai 400076, India}

\vspace{1em}

Nirpendra Singh\textsuperscript{\dag}\\
{\itshape Department of Physics, Khalifa University, Abu Dhabi-127788, United Arab Emirates (UAE) and\\
Research and Innovation Center for Graphene and 2D materials (RIC2D),\\
Khalifa University, Abu Dhabi, United Arab Emirates}
\end{center}

\vspace{2em}
Here, we provide further auxiliary details about the polyhedral structure of \textit{so}-MoS$_2$, trigonal pyramidal configuration of \textit{so}-SnS monolayer, phonon DOS of \textit{so} MoS$_2$ and \textit{so}-SnS, and their respective thermodynamic stability using molecular dynamics calcultions, calculated lattice thermal conductivity of MoS$_2$ and SnS in their hexagonal and \textit{so}-lattices, simulated elastic constants, bond length, bond stiffness and HSE06 band gaps.  

\begin{figure}[h]
\centering
\includegraphics[scale=0.03]{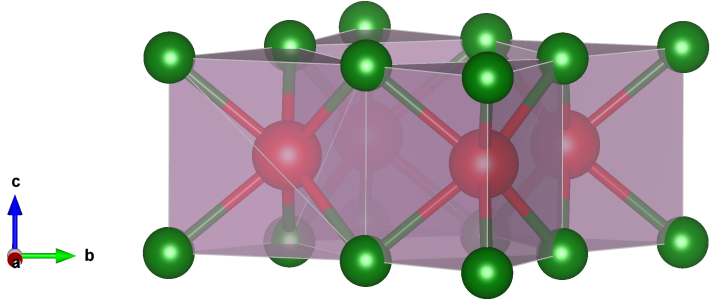}
\caption{The polyhedral structure of \textit{so}-MoS$_2$ monolayer.}
\label{polyhedra_MoS2}
\end{figure}

\begin{figure}[h]
\centering
\includegraphics[width=0.5\linewidth]{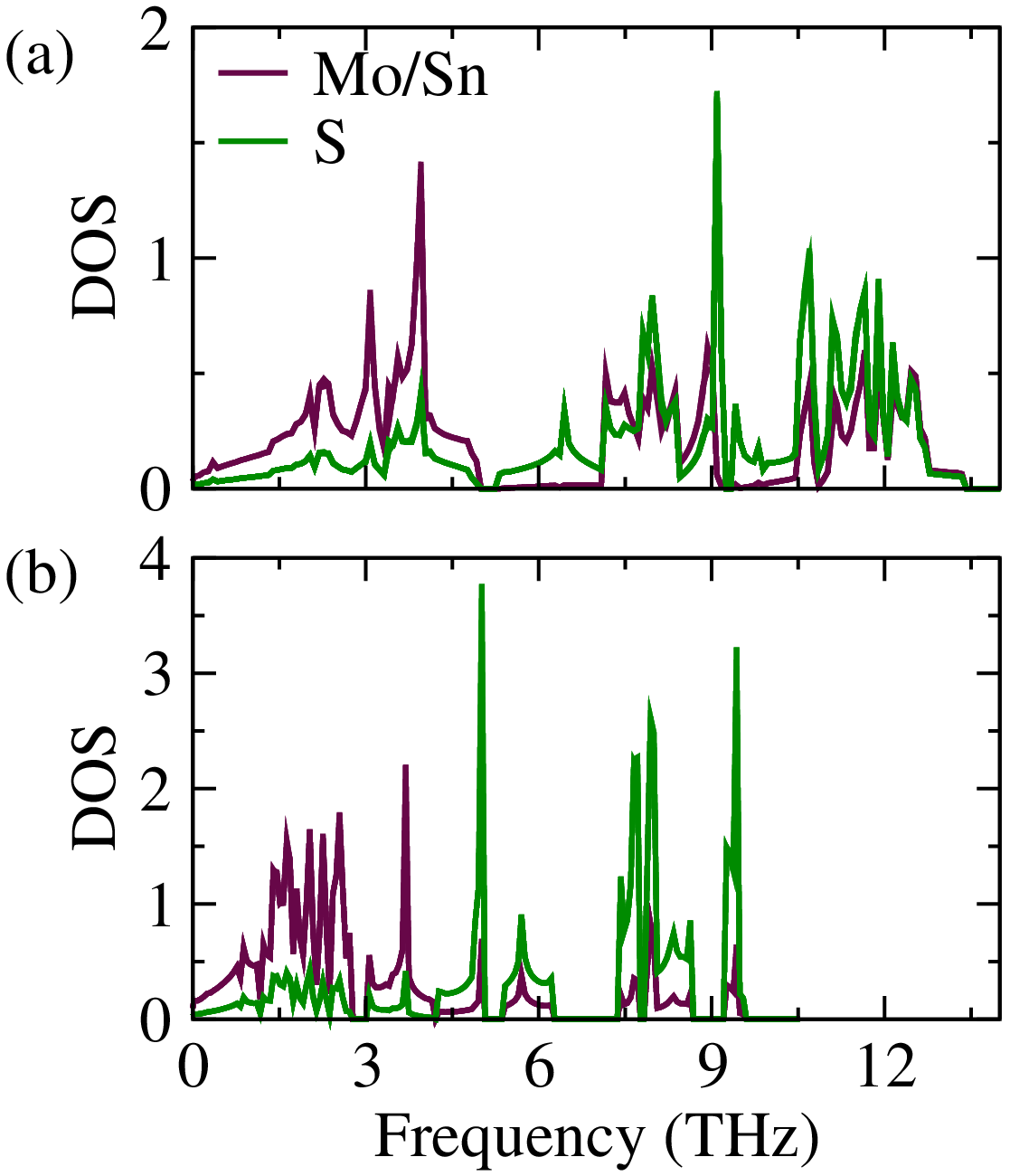} 
\caption{Calculated phonon density of states of (a) \textit{so}-MoS$_2$ and (b) \textit{so}-SnS monolayers.}
\label{phdos_so-MS2}
\end{figure}

\begin{table}[h]
\centering
\renewcommand{\arraystretch}{1.5}
\caption{Calculated bond length ($d_i$), bond stiffness ($k$), and HSE bandgap ($E_g$) of \textit{so} monolayers.}
\label{hse_SnX}

\begin{tabular}{||c|c|c|c|c|c||}
\hline\hline
\textbf{Monolayer} & \textbf{$d_1$} & \textbf{$k_1$} &
\textbf{$d_2$} & \textbf{$k_2$} & \textbf{$E_g$} \\
\hline
\textit{so}-MoS$_2$ & 2.45 & 3.59 & 2.37 & 23.19 & 0.09 \\
\textit{so}-SnS      & 2.53 & 0.24 & 2.67 & 0.42  & 3.00 \\
\hline\hline
\end{tabular}
\end{table}

\begin{table}[h]
\centering
\renewcommand{\arraystretch}{1.5}
\caption{Calculated elastic constants of studied \textit{so} monolayers.}
\label{elastic_so-MX}

\begin{tabular}{||c|c|c|c||}
\hline\hline
\textbf{Monolayer} &
\textbf{$C_{11}$ (N/m)} &
\textbf{$C_{12}$ (N/m)} &
\textbf{$C_{66}$ (N/m)} \\
\hline
\textit{so}-MoS$_2$ & 61.15 & 48.07 & 34.48 \\
\textit{so}-SnS      & 17.32 & 11.29 & 10.276 \\
\hline\hline
\end{tabular}
\end{table}

\end{document}